# PIONIC SUPERGIANT RADIOHALOS AS INTEGRAL RECORD OF PION EMISSION DURING NUCLEAR FISSION

D. B. Ion[1, 3], Reveica Ion-Mihai[2], M. L.D. Ion[2]
[1] *Institute for Physics and Nuclear Engineering, IFIN-HH, Bucharest Romania*
[2] *Bucharest University, Faculty of Physics, Bucharest, Romania*
[3] *Academy of Romanian Scientists*

*Abstract*: In this paper we presented a short review of radioactive halos as from the perspective of their interpretation as integral record in time of different kind of known or unknown radioactivities. A special attention is paid for the unified interpretation of the supergiant halos (SGH), discovered by Grady, Walker and Laemlein, as integral record of pion emission during fission.
*Key* words: pionic radioactivity, radiohalos, crystals, uranium

**Radioactive halos (Radiohalos).** It is well known that the pionic radioactivity was introduced by D.B. Ion, M. Ivascu and R. Ion-Mihai in Ref. [1]. Moreover supergiant halos as an integral record of pionic radioactivity was suggested in Ref. [2]. In this paper SGH (supergiant radioactive halos) are reviewed form the perspective of their unified interpretation as integral record of exotic decays such as spontaneous pion emission during fission.

The radioactive halos (see Ref. [3]) were first reported between 1880 and 1890 and their origin was a mystery until the discovery of radioactivity and its power of coloration. The radioactive halos are spherical, microscopic-sized discolorations in crystals. In cross sections on a microscope slide, they appear as a series of tiny concentric rings, usually surrounding a central core (see Fig.1). This central core is (or at least initially was) a radioactive inclusion in crystal. The alpha-particles, emitted from the radioactive inclusion during radioactive decay, damage the mineral and discolor it, with most of the damage occurring where the particle stops. How far this alpha-particle travels depends on its energy.

A large number of alpha particles are needed to form a visible radiohalo. It has been determined that to form a full dark radiohalo takes about 500 million to 1 billion alpha particles. This amounts to about 100 million years of decay at today's decay rates. But these radiohalos were formed under extreme temperatures of about 150 degrees Centigrade (or 302 degrees Fahrenheit). In addition, if the rock is reheated at a later time, the radiohalos will fade and disappear as the crystal atoms realign themselves and repair the crystal defects. Granites around the world contain dark radiohalos.

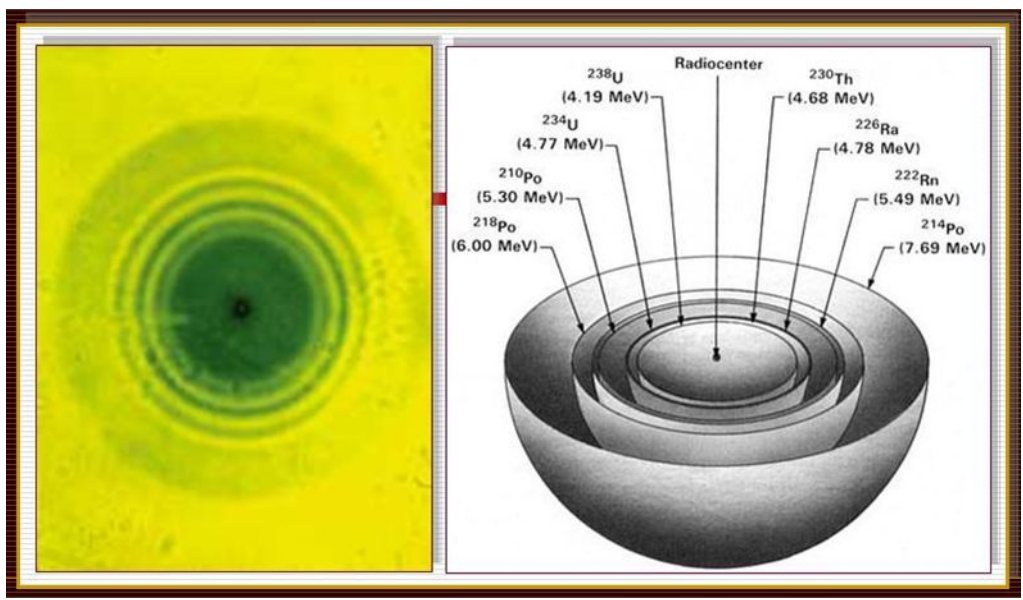

Fig.1: A fully-developed uranium radiohalo in biotite (left). An uranium halo comprises eight rings, but some rings are of similar size and cannot easily be distinguished. Color photos of uranium halos appear in the Radiohalo Catalogue [238]U Halo Cross Section (right). Idealized three-dimensional illustration of the [238]U-halo obtained by slicing the halo through the center. Each halo ring is identified by the appropriate isotope and its alpha energy in MeV.

Radiohalos are made up of either *uranium* or *polonium*, but there is a mystery surrounding these polonium radiohalos. The unstable polonium radiohalos have only short existences or decay rates. For example, polonium-218 has a 3-minute existence, polonium-210 has a 138-day existence and polonium-214 has a 164-microsecond existence.

So, $^{238}$U initiates a chain of steps which ends in the element lead (chemical symbol Pb). The $^{238}$U decay chain, as shown below (Fig.2), has some daughters which decay by emitting a beta (β) particle. The type of decay is shown by the symbols α and β. Since all the α-particles from a particular parent nucleus have the same energy and the particles are fired in all directions, a spherical shell of discoloration will be produced (see Fig.1b), appearing circular in cross-section. Hence, radioactive uranium $^{238}$U from inclusions generates multi-ringed halo because of its radioactive decay series:

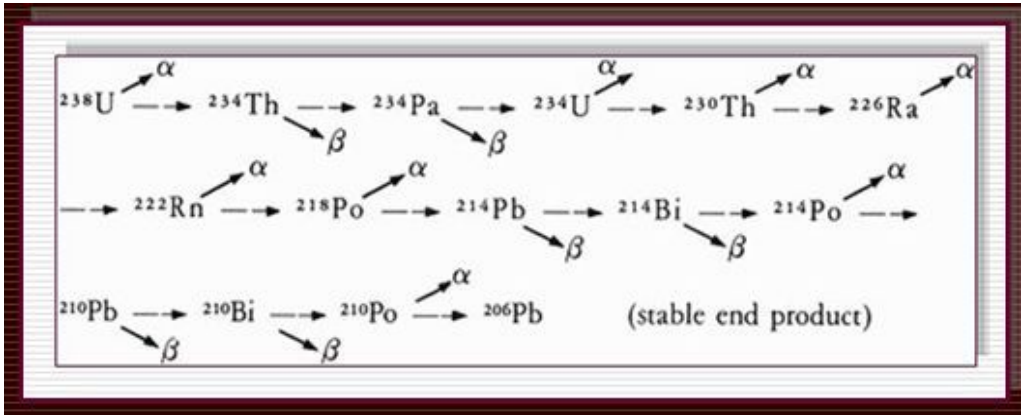

Fig. 2: The uranium-238 ($^{238}$U) series. Eight of the fifteen isotopes emit an alpha particle when they decay. This scheme shows that there are also six beta emitters in this chain, but their interaction with mica is insufficient to produce halo rings.

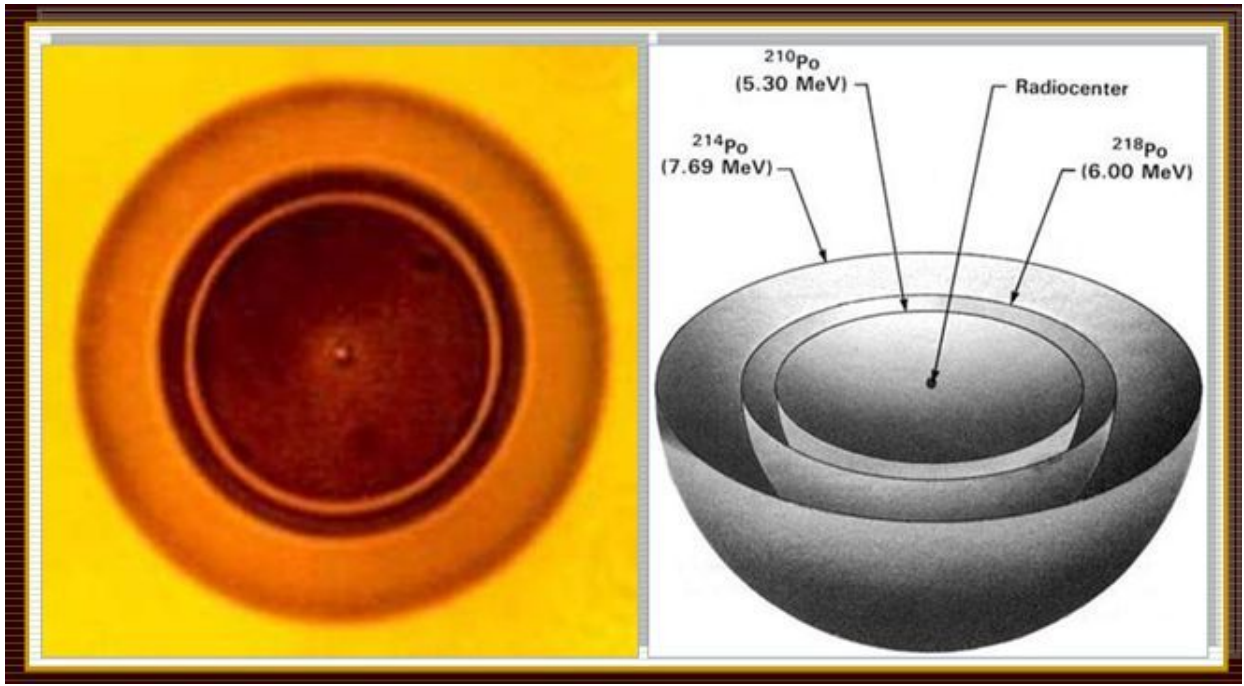

Fig. 3: (a) A fully Polonium-218 radiohalo in biotite (left). A Polonium halo comprises three rings, $^{218}$Po Halo Cross Section (right). Idealized three-dimensional illustration of the $^{238}$Po-halo obtained by slicing the halo through the center. Each halo ring is identified by the appropriate isotope and its alpha energy in MeV (see Ref. [4]).

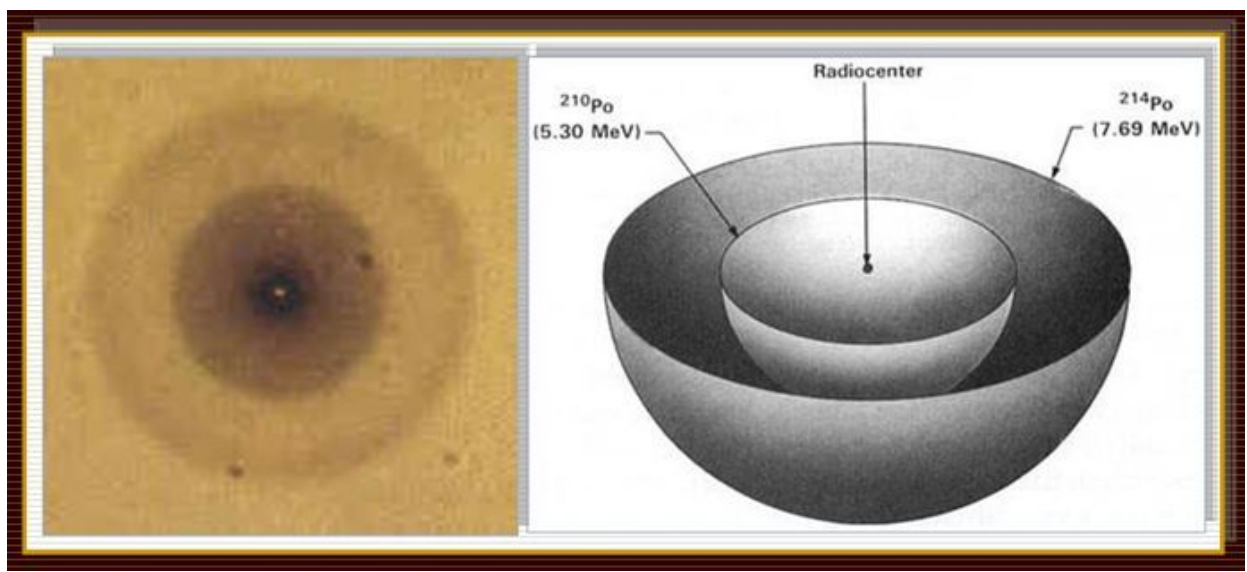

Fig. 3: (b) A fully Polonium-214 radiohalo in biotite (left). A Polonium halo comprises two rings, $^{214}$Po Halo Cross Section (right). Idealized three-dimensional illustration of the $^{214}$Po-halo obtained by slicing the halo through the center. Each halo ring is identified by the appropriate isotope and its alpha energy in MeV (see Ref. [4]).

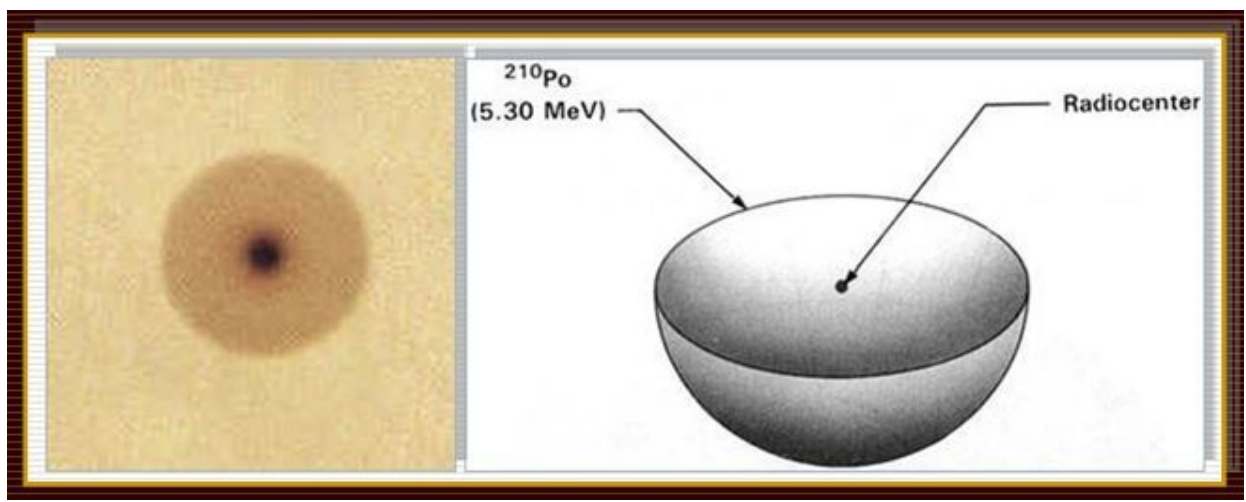

Fig. 3: (c) A fully Polonium-210 radiohalo in biotite (left). A Polonium halo comprises one rings, $^{210}$Po Halo Cross Section (right). Idealized three-dimensional illustration of the $^{210}$Po-halo obtained by slicing the halo through the center. Each halo ring is identified by the appropriate isotope and its alpha energy in MeV (see Ref. [4]).

**New kind of nuclear radioactivities.** Aside from their interest as attractive mineralogical oddities, the halos are of great interest for the nuclear physics because they are an integral record of radioactive nuclear decay in minerals. This integral record is detailed enough to allow estimation of the energy involved in the decay process and to identify the decaying nuclides through genetic decay chain. This latter possibility is particularly exciting because there are certain classes of halos, such as [3]: _dwarf halos_, _X-halos_, _giant halos_ and the _supergiant halos_ [4-5], which cannot be identified with the ring structure of the known alpha-emitters. Hence, barring the possibility of a non radioactive origin, these new variants of halos can be interpreted as evidences for hitherto undiscovered alpha-radionuclide, as well as, signals for the existence of new types of radioactivities.

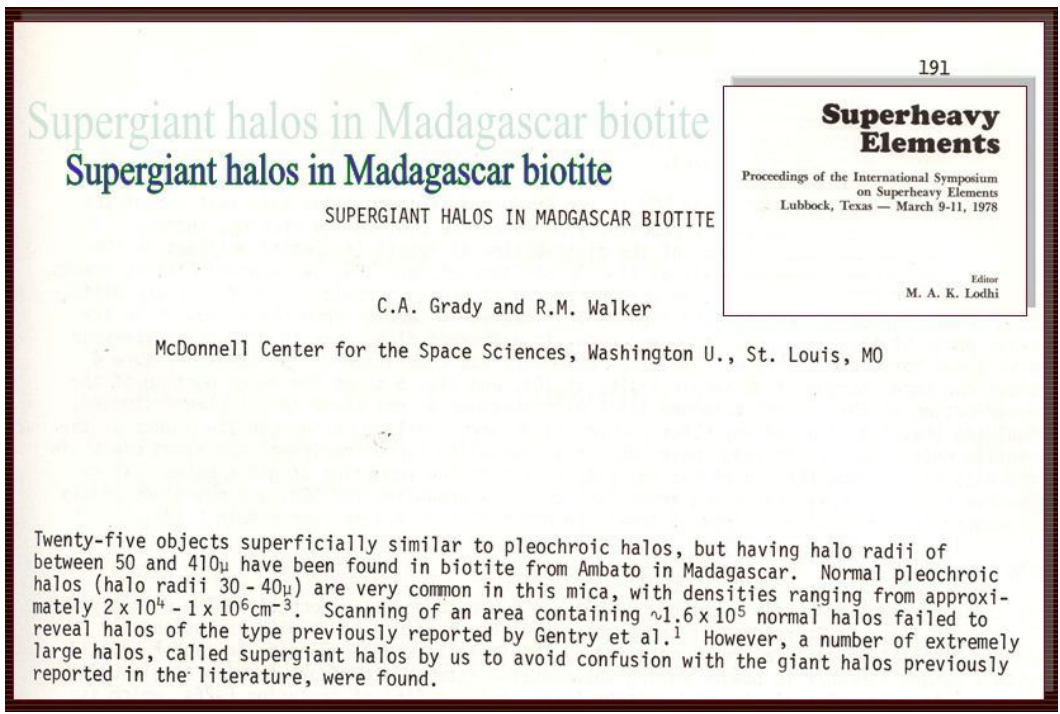

**Fig.4**: The abstract of the paper: *Supergiant Halos in Madagascar Biotite* (Ref. [8])

Therefore we must underline that in addition to the Laemmlein discovery [9], who found halos of rather diffuse boundaries with radii up to several thousand micrometers surrounding Thorium containing monazite inclusions in quartz, Grady and Walker [8] found twenty-five extremely large halos called supergiant halos (SGH) (see Fig.4). The SGH have the following essential characteristic features:

1. The SGH radii are between 50µm and 410µm. They have large oval to circular inclusions with radii of 20-52 µm.
2. SGH do not have sharp edges.
3. In thick sections (>30µm) SGH are surrounded by approximately circular brown regions having the same color in unpolarized light as the normal halos and exhibiting the same colors of pleochroism.
4. The coloration of SGH gradually fades as they move away from inclusion.
5. The SGH, with inclusions removed and etched with HF, show an abundance of fossil fission tracks in immediate proximity to the inclusion.
6. Three-dimensional etching of the SGH revealed a complex pattern of extended fossil fission tracks distribution throughout the halo.
7. The SGH inclusions were found to be alpha-active.

In the papers [2,5-8] we discussed a completely new possibility for a unified interpretation of the supergiant halos (SGH), namely, that SGH's are the integral records of the nuclear pionic radioactivity.

**Pionic Supergiant Radiohalos** [1,2,5-8]. The spontaneous pion emission accompanied by fission of heavy nuclei has been introduced in 1985 as a new of natural radioactivity, called *nuclear pionic radioactivity* (NPIR). The creation and emission of pions from the ground state of a nucleus is energetically possible only via two-body or many-body fission of the parent nuclei with Z larger than 80 (see Fig. 5)

The nuclear pionic radioactivity of a parent nucleus (A,Z) can be considered as an inclusive reaction of form:

$$(A,Z) \to \pi^{\pm,0} + X \tag{1}$$

where X denotes any configuration of final particles (fragments, light neutral and charged particles, etc.) which accompany emission process. The inclusive NPIR is in fact a sum of all exclusive nuclear reactions allowed by the conservation laws in which a pion can be emitted by a nucleus from its ground state. The important exclusive reactions which give the essential contribution to the inclusive NPIR (1) are the spontaneous pion emission accompanied by two body fission:

$$(A,Z) \to \pi^{\pm,0} + (A_1, Z_1) + (A_2, Z_2) \quad \text{where} \quad A = A_1 + A_2 \quad \text{and} \quad Z = Z_1 + Z_2 + Z_\pi \quad (2)$$

Hence, the NPIR is an extremely complex coherent reaction in which we are dealing with a spontaneous pion emission accompanied by a rearrangement of the parent nucleus in two or many final nuclei. Charged pions as well as neutral pions can be emitted. A fission like model for the NPIR is regarded as a first stage in the development of an approximate theory of this new phenomenon (see for example the papers [2,5-8], that takes into account the essential degree of freedom of the system $\pi^{\pm,0} - fissility$, $X_{\pi F}$. Physical regions from the (A, Z)-plane from which the nuclei are able to emit pions, spontaneously, are presented in Fig. 5

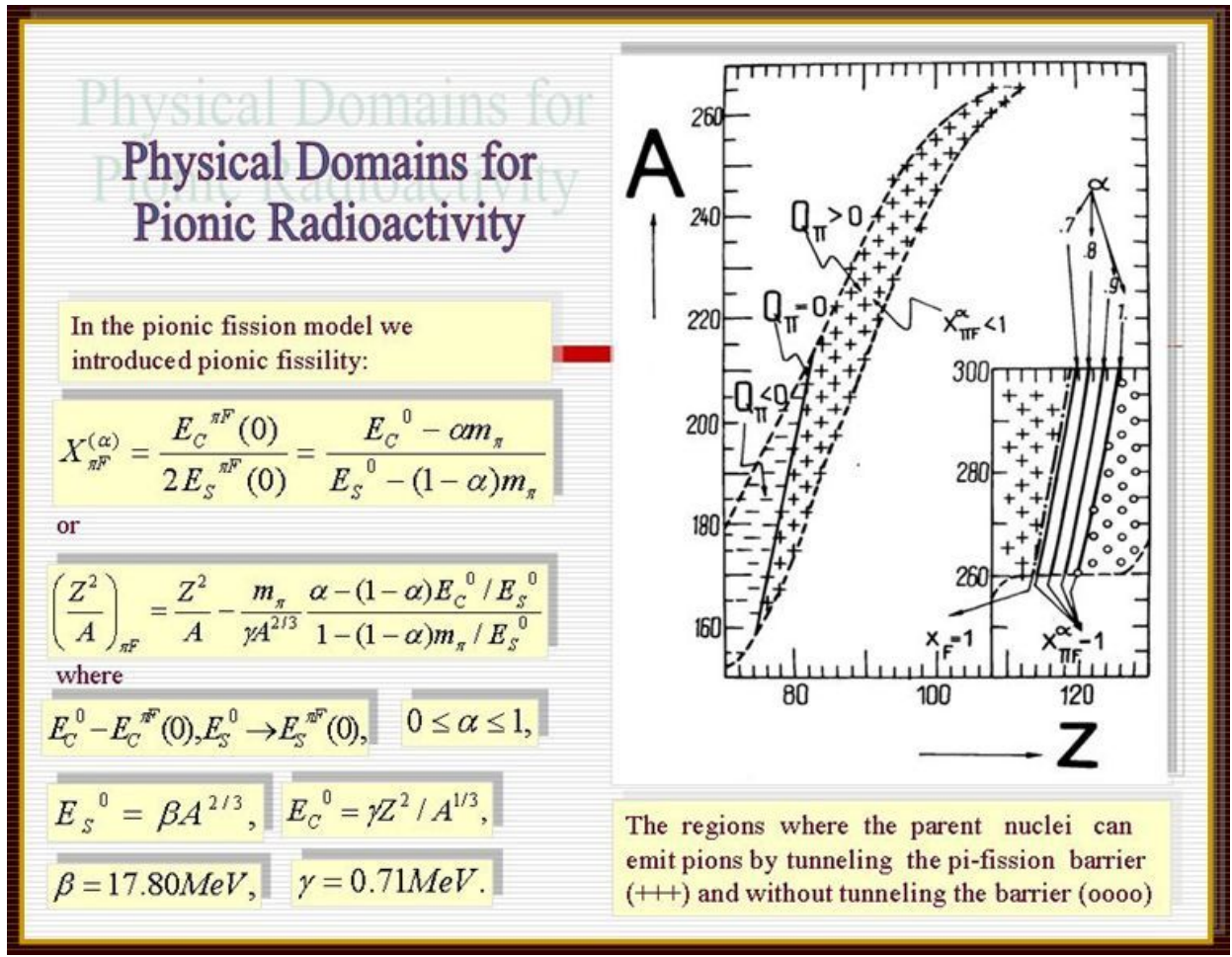

Fig.5: Physical regions for the pionic radioactivity [2, 5-8]

Therefore, we must show that a nuclear reaction (2) is able to produce the supergiant halos discovered by Grady and Walker [8] and Laemmlein [9].

Now, in order for a "*discolored*" region in a transparent mineral to be truly a *pionic radioactive halo* (PIRH), it must satisfy the following rules:

**R1**: It must have an inclusion that is, or at least one time was, radioactive. The dimension of the PIRH inclusion must be sufficiently large in order to satisfy the dose rule R2.

**R2**: The dimension of a radioactive halo is given by that particle emission process with largest range with the condition that its dose satisfies the coloration threshold condition.

**R3**: In the nuclear reactions (1) from the PIRH-inclusions, the $\pi^-$-yields should be about two orders of magnitude larger than the $\pi^+$-yields (see Ref.[2,5-7])

*Definition of Pionic Radiohalos.* A radioactive halo is called pionic radiohalo (PIRH) if its inclusion (R1) contains, or at one time contained, a parent nucleus pionic radioactive (1) in such a concentration that produced a pionic dose satisfying the coloration threshold condition (R2). The PIRH by definition must be the integral record in time of the nuclear pionic radioactivity in some minerals such as biotite, fluorite, cordierite, etc.

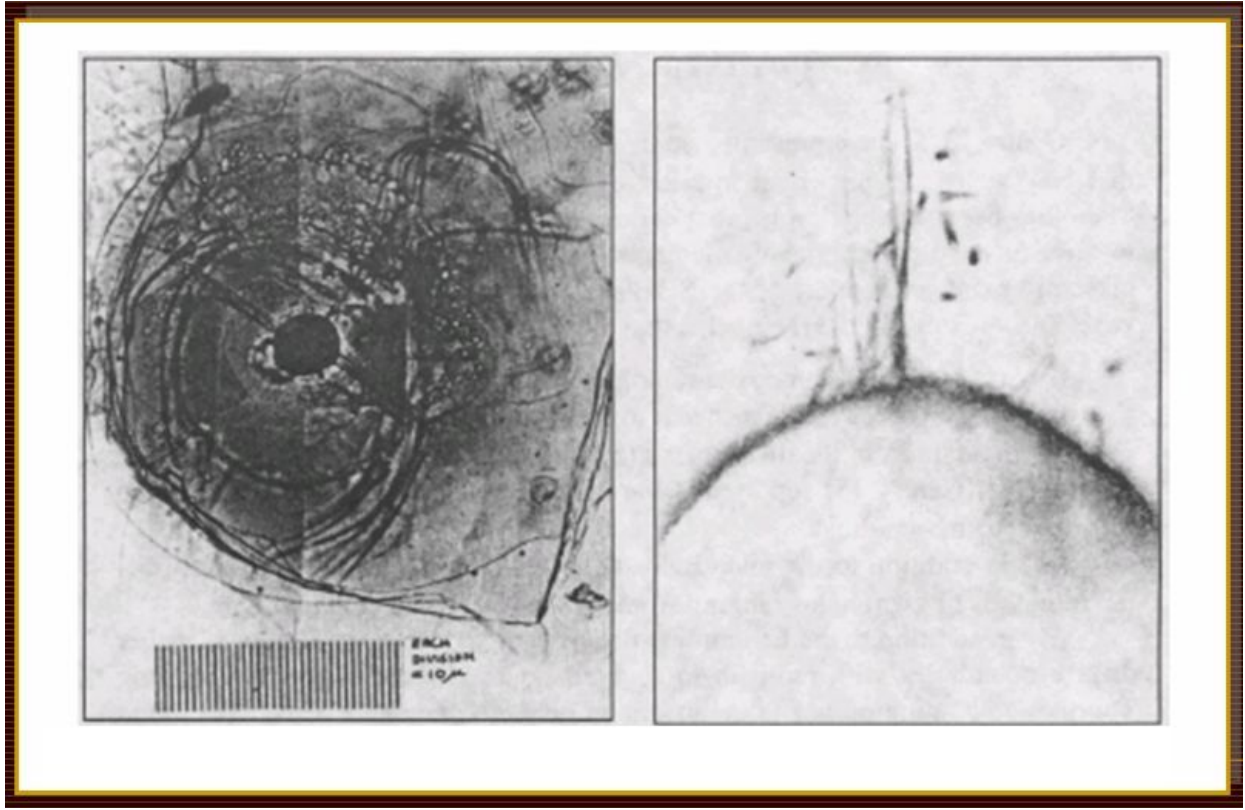

Fig. 6a: A photomosaic of a part of one supergiant halos (SGH) F-12 discovered by Grady and Walker [9](left). The essential characteristic features of this SGH are well reproduced by those of PIRH supergiant halos (see the text in [2,5-7]); Fig. 6b: A photoimage of the mica surface showing fossil tracks next to inclusion pit (right).

According to R3-rule (see Ref. [2]), we expect that two types of PIRH are possible to be observed in nature. These will be the following:

- PIRH(-), when only $\pi^-$ satisfy the dose rule R2;
- SPIRH, which is defined just as a superposition of PIRH(-) and PIRH(+), when both $\pi^-$ and $\pi^+$ satisfy the dose rule R2.

PIRH(-) and SPIRH have the same signatures, except a big difference (about 700-800μm in radius, in favor of SPIRH) in dimensions. This difference is given by the fact that $\pi^+$ mesons are mainly decaying via

$$\pi^+ \to \mu^+ + \bar{v}_\mu$$
$$\downarrow$$
$$e^+ + v_e + v_\mu \qquad (3)$$

while the $\pi^-$ are stopped in mineral (before their natural decay occurs) producing the fission of the mineral nuclides in two or more fragments. Knowing that electrons do not contribute to the radiohalos range, the radius of the SPIRH would be given by the range of $\pi^+$ to which is added the range of its decay product $\mu^+$, while the radius of PIRH(-) is given only by the range of $\pi^-$.

The PIRH signatures are as follows [2]:
♦ PIRH is a pionic radiohalo of supergiant dimensions which possesses a large inclusion, sufficient to satisfy the rule R2. The supergiant PIRH dimensions are essentially determined by the high ranges of $\pi$ mesons as well as of their decay product, $\mu$ leptons (see Fig. 7)

♦ The PIRH do not have sharps edges and their coloration gradually fades as moves away from inclusion;

♦ The PIRH must show an abundance of fossil fission tracks in the immediate proximity to the inclusion. This signature is given just by the fission fragments of the pionic emitters (see reaction (1)) from the PIRH-inclusion.

♦ The PIRH inclusions, by definition, can be α-active. As a consequence of the α-activity of the nuclei from the PIRH inclusions they must be surrounded by approximately circular regions having the same color as normal halos and exhibiting the same colors of pleocroism;

♦ Extended fossil fission tracks distribution through the PIRH must be also present as a consequence of the fission of the mineral nuclides produced by the stopping $\pi^-$ mesons in mineral.

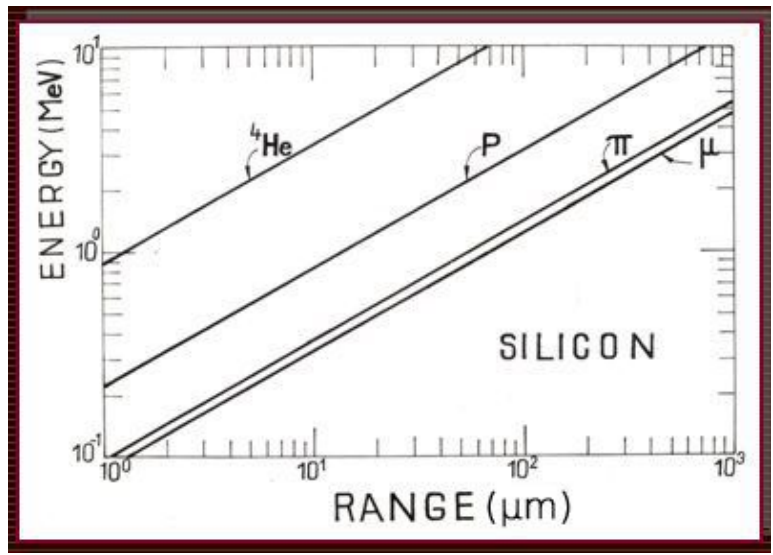

Fig. 7. Energy-range diagrams for muons, pions, protons and alpha particles in silicon
(for biotite, fluorite and cordierite see Fig. 4 in Ref.[2])

Next, comparing the PIRH signatures with the above essential characteristic features (1-7) of the SGH, it is easy to see that the PIRH(-) can be identified, with a surprising high accuracy, as being the SGH discovered by Grady and Walker [8] in biotite.

We note that a special investigation and the discovery of the Laemmlein-like [9] halos are necessary since these SGH with dimensions around 1-1.5 mm can be identified as being the SPIRH halos produced by emission of $\pi^+$ mesons during fission of the parent nuclei from the halos inclusion.